\documentclass[aip,jap,preprint]{revtex4-1}

\usepackage{amsmath,epsfig,amsfonts,graphicx,dcolumn,bm}

\begin{document}

\title[Tunable Vibrational Band Gaps in One-Dimensional Diatomic Granular Crystals with Three-Particle Unit Cells]{Tunable Vibrational Band Gaps in One-Dimensional Diatomic Granular Crystals with Three-Particle Unit Cells}

\author{N. Boechler}
\author{J. Yang}%
\author{G. Theocharis}
\affiliation{ 
Graduate Aerospace Laboratories (GALCIT), California Institute of Technology, Pasadena, CA 91125, USA
}
\author{P. G. Kevrekidis}%
\affiliation{ 
Department of Mathematics and Statistics, University of Massachusetts, Amherst MA 01003-4515, USA 
}
\author{C. Daraio}%
\affiliation{ 
Graduate Aerospace Laboratories (GALCIT), California Institute of Technology, Pasadena, CA 91125, USA
}%

\date{\today}


\begin{abstract}
We investigate the tunable vibration filtering properties of one-dimensional diatomic granular crystals composed of arrays of stainless steel spheres and cylinders interacting via Hertzian contact.  The arrays consist of periodically repeated three-particle unit cells (steel-cylinder-sphere) in which the length of the cylinder is varied systematically. We apply static compression to linearize the dynamic response of the crystals and characterize their linear frequency spectrum. We find good agreement between theoretical dispersion relation analysis (for infinite systems), state-space analysis (for finite systems), and experiments. We report the observation of up to three distinct pass bands and two finite band gaps and show their tunability for variations in cylinder length and static compression. 
\end{abstract}


\maketitle

\section{\label{sec:level1}Introduction}

The presence of band gaps, a characteristic of wave propagation in periodic structures, has been studied in a wide
array of settings involving phononic crystals, photonics, and plasmonics 
\cite{Brillouin,Duang05,Soukoulis08,kivshar03,Brongersam}. 
Materials exhibiting band gaps
are of particular interest as they forbid and allow the propagation of waves in selected frequency ranges (pass and
stop bands), and in the case of elastic wave propagation (in composites or multilayered structures) have previously been proposed for use in acoustic filters, vibration isolation applications, and rectification of acoustic energy flux \cite{Sala,Sigalas,Fok,Liang}.

Chains composed of elastic particles in close contact with each other, or ``granular crystals", have gained much attention with
respect to elastic wave propagation in nonlinear media. The nonlinearity in granular crystals results from the Hertzian contact between
two elastic spherical (or spherical and cylindrical) particles in compression and from a zero tensile strength. The contact stiffness is defined by the geometry
and material properties of the particles in contact \cite{Johnson}. In this type of system the level of nonlinearity present can be controlled by the amount of static compression applied to the chain, resulting in a dynamic response which encompasses the linear, weakly nonlinear, and strongly nonlinear regimes \cite{nesterenko1, dar06}. This simple means of controlling their dynamic response has made granular cystals a perfect test bed for the study of nonlinear phenomena, including the emergence of coherent structures such as solitary waves \cite{nesterenko1,Coste97}, discrete breathers \cite{Boechler10,sen09}, shock waves \cite{daraio09}, and linear/nonlinear 
defect modes \cite{Melo09,Theo09}. Additionally, granular crystals have been shown to be useful in application to engineering endeavors including shock and energy absorbing layers \cite{dar06,hong05,fernando,doney06}, actuating devices \cite{dev08}, acoustic lenses \cite{Spadoni} and sound scramblers \cite{dar05,dar05b}.

Previous studies involving statically compressed granular crystals composed of one-dimensional (1D)
periodic (monoatomic and diatomic with a two particle unit cell) arrays of glued \cite{Hennion07}, welded \cite{Hennion05}, and elastically compressed spherical 
particles \cite{Billy00,Boechler09,Boechler10,Herbold} have been shown to exhibit tunable frequency vibrational band gaps. In this manuscript we study  statically compressed 1D diatomic granular crystals  composed of periodic arrays of stainless steel sphere-cylinder-sphere unit cells. We employ theoretical models to estimate the dispersion relation of the crystals, we numerically validate their dynamic response using state-space analysis and verify experimentally the crystal's acoustic transmission spectrum. 
For such  configurations we experimentally report the presence of a third distinct pass band and a second finite band gap. We show tunability and
customization of the response for variation of the cylinder length and static compression. \newline

\section{\label{sec:level1}Experimental setup}

\begin{figure}[h]
\begin{center}
\includegraphics[width=7.2cm,height=3.6cm]{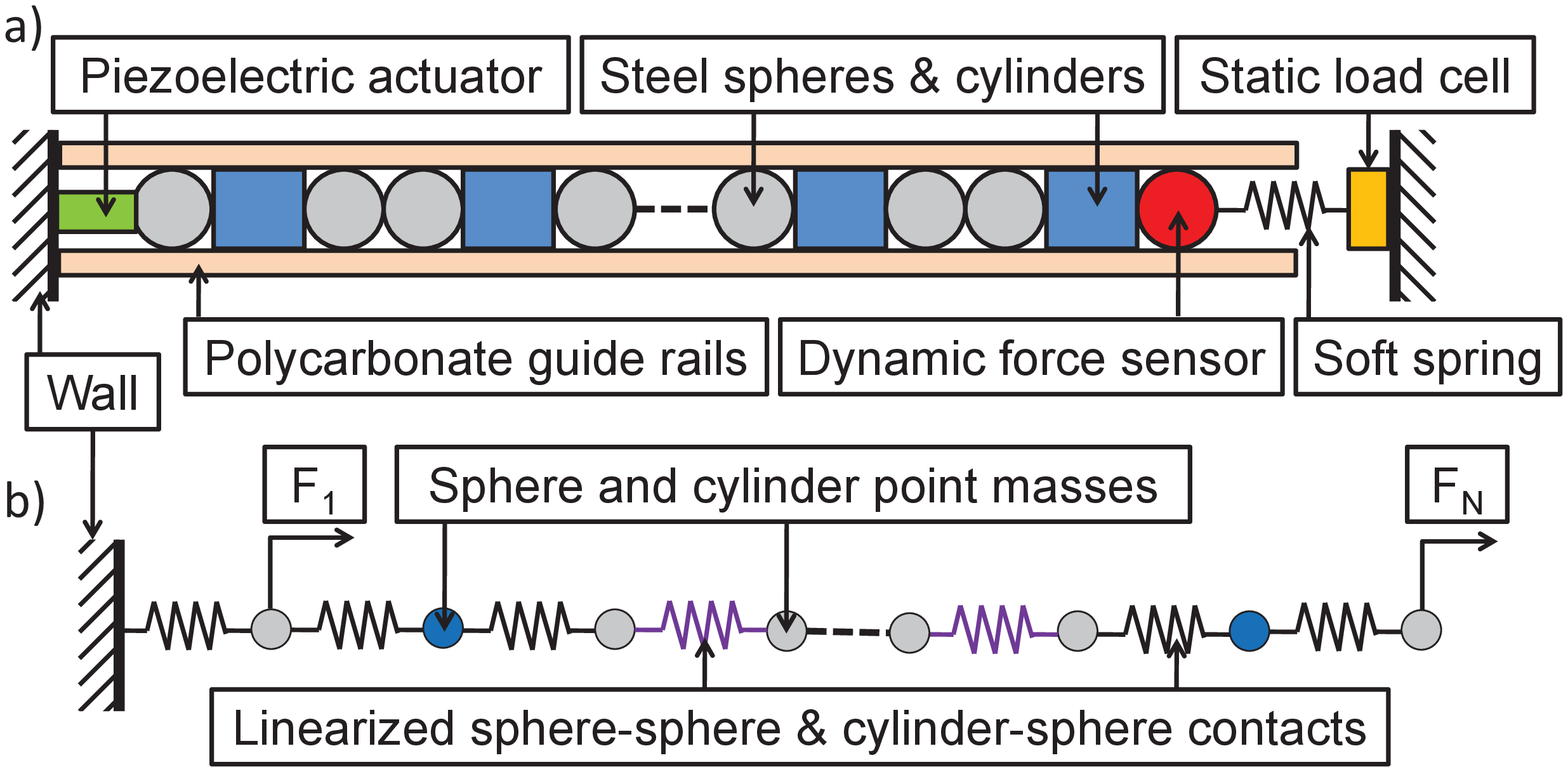}
\end{center}
\caption{\label{Fig1}[Color online] a) Schematic of experimental setup. b) Schematic of the linearized model of the experimental setup.}
\end{figure}

We assemble five different 1D diatomic granular crystals composed of three-particle, sphere-cylinder-sphere, repeating unit cells as shown in Fig.~\ref{Fig1}(a). The chains are 21 particles (7 unit cells) long. The particles (spheres and cylinders) are made from $440$C stainless steel, with radius $R=9.53$~mm, elastic modulus $E=200$~GPa, and Poisson's ratio $\nu=0.3$ ~\cite{ElasticProperties}. Each of the five chains is assembled with cylinders of a different length, $L=[9.4,~12.5,~15.8,~18.7,~21.9]$~mm. The mass of the spherical particles is measured to be $m=27.8$~g and the mass of the cylindrical particles is measured to be $M=[20.5,~27.3,~34.1,~40.7,~47.8]$~g for each of the corresponding cylinder lengths. 

We align the spheres and cylinders in a horizontal 1D configuration using a containment structure of four polycarbonate rods ($12.7$~mm diameter). We hold the polycarbonate rods in place with polycarbonate guide plates spaced at intervals of 1 unit cell. We apply low amplitude broadband noise to the granular crystals using a piezoelectric actuator mounted on a steel cube of height $88.9$~mm which is fixed to the table. We visualize the evolution of the force-time history of the propagating excitations using a calibrated dynamic force sensor. The force sensor is composed of a piezo-electric disk embedded with epoxy inside two halves of a $R=9.53$~mm, $316$ stainless steel sphere (of elastic modulus $193$~GPa, and a Poisson ratio of $0.3$~~\cite{ElasticProperties}). The sensor is constructed so as to approximate the mass, shape, and contact properties of the spherical particles in the rest of the crystal~\cite{NesterenkoSensor,dar05,dar05b,dar06}. The assembled force sensor is calibrated against a commercial dynamic force sensor, and has a measured total mass and resonant frequency of $28.0$~g and $80$~kHz, respectively. We insert the dynamic force sensor in place of the last particle, located at the opposite end of the crystal from the actuator. We condition its output with a $30$~kHz cutoff 8-pole butterworth low pass filter and voltage amplifier. 

At the opposite end of the crystal with respect to the piezoelectric actuator, we apply a static compressive force, $F_0$, using a soft (compared to the contact stiffness of the particles) stainless steel linear compression spring (stiffness $1.24$~kN/m). In this case we can approximate this boundary as a free boundary. The static compressive force applied to the chain is adjusted by positioning, and fixing to the table, a movable steel cube of height $76.2$~mm so that the soft linear spring is compresssed. The resulting applied static load is measured with a static load cell placed in between the steel cube and the spring.   \newline


\section{\label{sec:level1}Theoretical setup}

\subsection{\label{sec:level11}Dispersion Relation}

We model a 1D diatomic crystal composed of $n$ sphere-cylinder-sphere unit cells (and $N$ particles) as a chain of nonlinear oscillators~\cite{nesterenko1}:
\begin{equation}
\begin{aligned}
m_{l}\ddot{u}_l&=\alpha_{l-1,l}[\delta_{l-1,l}+u_{l-1} - u_{l}]_{+}^{p} \\
							 &- \alpha_{l,l+1}[\delta_{l,l+1}+u_{l} - u_{l+1}]_{+}^{p}, \\
\end{aligned}
\label{model}
\end{equation}
where $[Y]_+$ denotes the positive part of $Y$; the bracket takes the value $Y$ if $Y>0$, and 0 if $Y\leq0$. This represents the tensionless characteristic of our system: when adjacent particles are not in contact, there is no force between them. The above model assumes that the particles act as point masses. This is valid as long as the frequencies of the applied vibrations are much lower than the frequencies of the natural vibrational modes of the individual particles~\cite{Hennion02}. Here, $u_l$ is the displacement of the $l$th particle around the static equilibrium, $\delta_{l-1,l}$ is the static overlap between the $(l-1)$th and the $l$th particles, and $m_l$ is the mass of the $l$th particle (where $l$ is the index of the $l$th particle in the chain counted from the piezoelectric actuator end, and $l \in \{1,\cdots,3n\}$). As per Hertz's contact law, the coefficients $\alpha$ depend on the geometry and material properties of the adjacent particles and on the exponent $p$ (here $p=3/2$)~~\cite{Johnson}. Here, in the case of the sphere-cylinder-sphere unit cell, we need to account for two different values of the contact coefficients $\alpha$, corresponding to the sphere-cylinder and the sphere-sphere contacts, where:  
\begin{equation}
\begin{aligned}
\alpha_{sphere,cylinder}=\alpha_{cylinder,sphere}=A_{1}=\frac{2E\sqrt{R}}{3(1-\nu^2)},
\end{aligned}
\label{Hertz1}
\end{equation}
\begin{equation}
\begin{aligned}
\alpha_{sphere,sphere}=A_{2}=\frac{E\sqrt{2R}}{3(1-\nu^2)}.
\end{aligned}
\label{Hertz2}
\end{equation}

For this case, it can be seen that $A_{1}=\sqrt{2}A_{2}$. Furthermore, for Hertzian contacts, under a static load $F_0$, we can define the static overlap for the sphere-cylinder contact as $\delta_{sphere,cylinder}=\delta_{cylinder,sphere}=(F_0/A_{1})^{2/3}$, and for 
the sphere-sphere contact as $\delta_{sphere,sphere}=(F_0/A_{2})^{2/3}$~~\cite{Johnson,nesterenko1}. Considering small amplitude dynamic displacements as compared to the static overlap, one can linearize the equations of motion (Eq.~\ref{model}).
For the studied sphere-cylinder-sphere unit cell the particles' linearized equations of motion are:  
\begin{equation}
\begin{aligned}
m\ddot{u}_{3j-2}&=\beta_2[u_{3j-3} - u_{3j-2}] - \beta_1[u_{3j-2} - u_{3j-1}], \\
M\ddot{u}_{3j-1}&=\beta_1[u_{3j-2} - u_{3j-1}] - \beta_1[u_{3j-1} - u_{3j}], \\
m\ddot{u}_{3j}&=\beta_1[u_{3j-1} - u_{3j}] - \beta_2[u_{3j} - u_{3j+1}], \\
\label{linearizedEOM}
\end{aligned}
\end{equation}
where $j$ is the number of the $j$th unit cell ($j \in \{1,\cdots,n\}$), $m$ is the mass of a spherical particle, $M$ is the mass of a cylindrical particle, $\beta_1=\frac{3}{2}A_1^{2/3}F_0^{1/3}$ is the linearized stiffness between a spherical and cylindrical particle, and $\beta_2=\frac{3}{2}A_2^{2/3}F_0^{1/3}$ is the linearized stiffness between two spherical particles. The dispersion relation for a diatomic (two particle unit cell) granular crystal is known to contain two branches (\emph{acoustic} and \emph{optical})~\cite{Herbold}. Here we use a similar procedure to calculate the dispersion relation for a diatomic crystal with a three particle unit cell. 

\begin{figure}[h]
\begin{center}
\includegraphics[width=9cm,height=5cm]{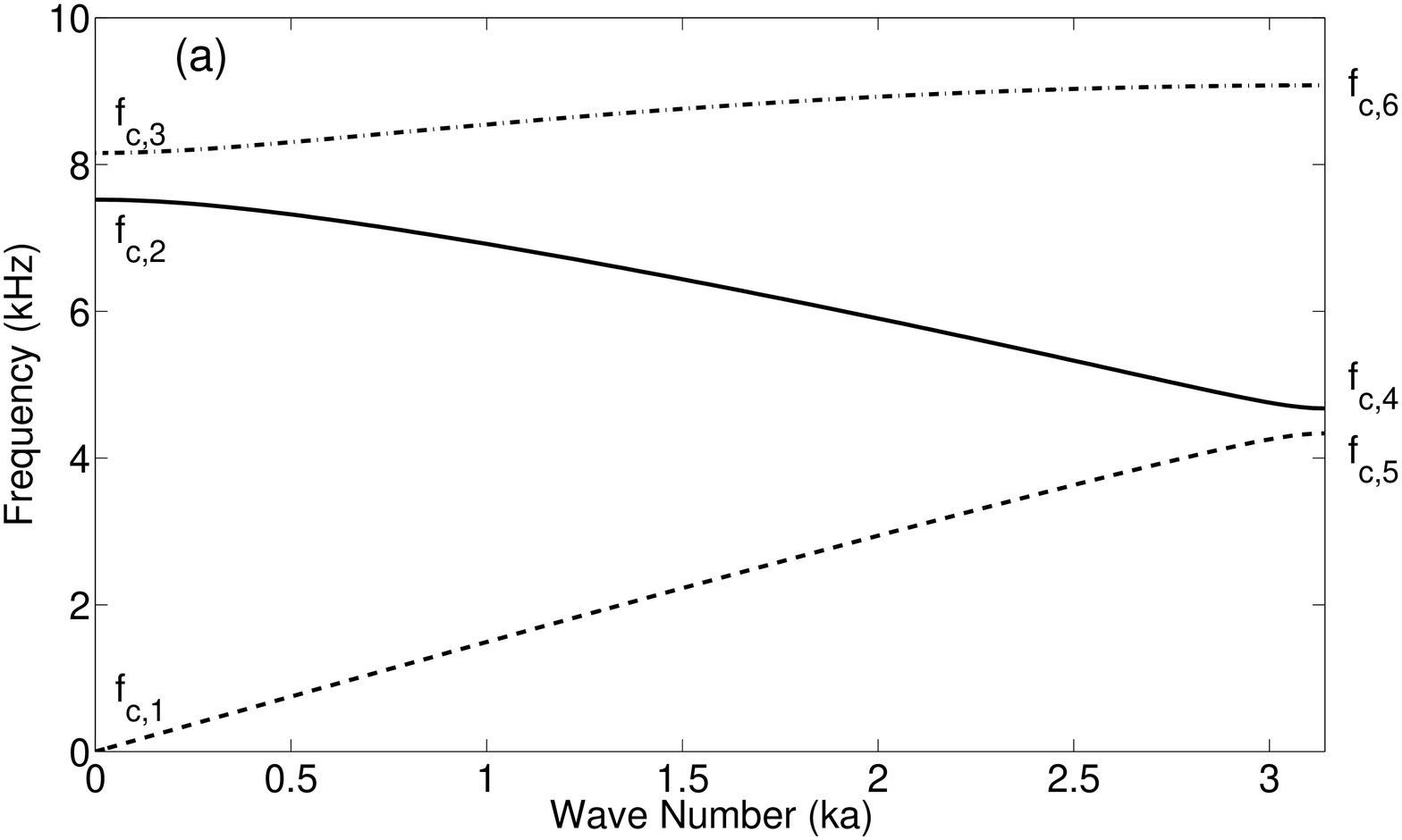}
\includegraphics[width=9cm,height=5cm]{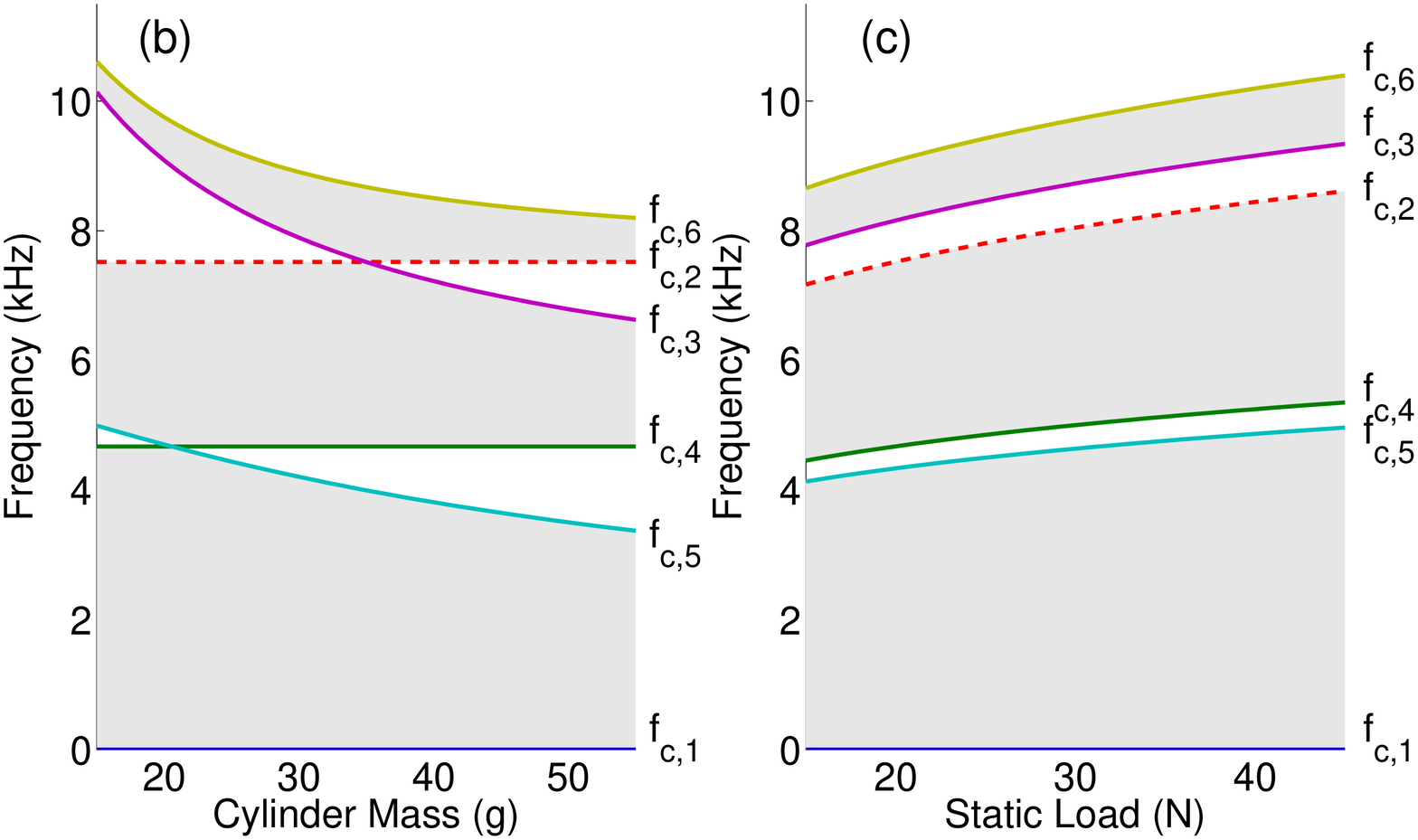}
\end{center} 
\caption{\label{Fig2} a) Dispersion relation for the described sphere-cylinder-sphere granular crystal with cylinder length $L=12.5$~mm ($M=27.3$~g) subject to a $F_0=20$~N static load. The acoustic branch is the dashed line, the lower optical branch is the solid line, and the upper optical branch is the dash-dotted line. Cutoff frequencies for granular crystals corresponding to our experimental configuration (b) varying the length L (and thus mass) of the cylinder with fixed $F_0$=20 N static compression, and (c) varing the static compression ($F_0=[20,~25,~30,~35,~40]$~N) with fixed $L=12.5$~mm cylinder length ($M=27.3$~g). Solid lines represent the six cutoff frequency solutions. $f_{c,2}$ is dashed to clarify the nature of the intersection with $f_{c,3}$. Shaded areas are the propagating bands.
}
\end{figure}

We substitute the following traveling wave solutions where $k$ is the wave number, $\omega$ is the angular frequency, and $a=L+4R-2\delta_{sphere,cylinder}-\delta_{sphere,sphere}$ is the equilibrium length of the sphere-cylinder-sphere unit cell: 
\begin{equation}
\begin{aligned}
{u}_{3j-2}&=Ue^{i(k a j+\omega t)}, \\
{u}_{3j-1}&=Ve^{i(k a j+\omega t)}, \\
{u}_{3j}&=We^{i(k a j+\omega t)}, \\
\label{travelingwave}
\end{aligned}
\end{equation}
into Eqs.~(\ref{linearizedEOM}). $U$, $V$, and $W$ are the wave amplitudes, and are constructed complex so as to contain both the amplitude and phase difference for each particle within the unit cell. Solving for a nontrivial solution we obtain the following dispersion relation:
\begin{equation}
\begin{aligned}
0=&-2{\beta_1}^2 \beta_2+\beta_1 (\beta_1+2\beta_2)(2m+M)\omega^2 \\
&-2m(\beta_2 M+\beta_1 (m+M)) \omega ^4 \\
&+m^2 M \omega ^6+2 {\beta_1}^2 \beta_2 \mathrm{cos}(ak).\\
\end{aligned}
\label{dispersionrelation}
\end{equation}
In Fig.~\ref{Fig2} (a), we plot the dispersion relation (Eq.~\ref{dispersionrelation}) for the previously described sphere-cylinder-sphere unit cell granular crystal with cylinder length $L=12.5$~mm ($M=27.3$~g) subject to an $F_0=20$~N static load. Three bands of solutions (or propagating frequencies) can be seen; the lowest in frequency being the acoustic band, followed by lower and upper optical bands. Frequencies in between these bands are said to lie in a band gap (or forbidden band). Waves at these frequencies are evanescent, decay exponentially, and cannot propagate throughout the crystal~\cite{Brillouin}. 

If we solve the dispersion relation, Eq.~(\ref{dispersionrelation}), for when $k=\frac{\pi}{a}$ and $k=0$ we obtain the following cutoff frequencies:  
\begin{equation}
\begin{aligned}
f_{c,1}^2&=0,\\
f_{c,2}^2&=\frac{\beta _1+2 \beta _2}{4\pi ^2m}, \\
f_{c,3}^2&=\frac{\beta _1(2m+M)}{4\pi ^2m M}, \\
f_{c,4}^2&=\frac{\beta _1}{4\pi ^2m}, \\
f_{c,5}^2&=\frac{ \beta _1 (2m+M)+2 \beta _2 M}{8\pi ^2 m M}\\
				&-\frac{\sqrt{-16 \beta _1 \beta _2 m M+\left(2 \beta _1 m+\beta _1 M+2 \beta _2 M\right){}^2}}{8\pi ^2 m M}, \\
f_{c,6}^2&=\frac{ \beta _1 (2m+M)+2 \beta _2 M}{8\pi ^2 m M}\\
				&+\frac{\sqrt{-16 \beta _1 \beta _2 m M+\left(2 \beta _1 m+\beta _1 M+2 \beta _2 M\right){}^2}}{8\pi ^2 m M}. \\
\end{aligned}
\label{cutofffrequencies}
\end{equation}
$f_{c,1}$, $f_{c,2}$, and $f_{c,3}$ correspond to $k=0$ and $f_{c,4}$, $f_{c,5}$, and $f_{c,6}$ to $k=\frac{\pi}{a}$. In Fig.~\ref{Fig2} (a), we label the six cutoff frequencies (Eqs.~\ref{cutofffrequencies}) for the previously described granular crystal with cylinder length $L=12.5$~mm ($M=27.3$~g) subject to a $F_0=20$~N static load.

From Eqs.~\ref{cutofffrequencies}, it can be seen that the cutoff frequencies are tunable through the variation of particle masses $m$ and $M$, and the linearized stiffnesses $\beta_1$ and $\beta_2$ (thus tunable with changes in geometry, and static compression $F_0$). In Fig. 2(b) we plot the cutoff frequencies in Eqs.~(\ref{cutofffrequencies}) as a function of cylinder length for fixed $F_0$=20~N static compression, and in Fig. 2(c) as a function of
static compression ($F_0=[20,~25,~30,~35,~40]$~N) for fixed cylinder length $L=12.5$~mm ($M=27.3$~g). The lines represent the cutoff frequency solutions ($f_{c,2}$ is dashed to clarify the nature of the intersection with $f_{c,3}$, and the shaded areas are the pass bands). It can be seen that within our frequency range of interest, two of the cutoff frequency solutions 
coincide at specific cylinder lengths. The intersection between $f_{c,4}$ and $f_{c,5}$ can be found to occur at $M/m=\frac{\beta_1}{\beta_2}$ and the intersection between $f_{c,2}$ and $f_{c,3}$ at $M/m=(2-\frac{\beta_1}{\beta_2})$. Notice, however, that aside from these special parameter
values where the above intersections occur, the spectrum preserves the three pass bands with two associated finite bandgaps between them.
\newline

\subsection{\label{sec:level2}State-space approach}

In addition to the dispersion relation previously calculated for an infinite system, we study the finite linearized system corresponding to our experimental setup as shown in Fig.~\ref{Fig1}(b). We model the actuator boundary of our system as a fixed 440C steel wall. We model the other end of the chain as a free boundary, as the stiffness of the spring used for static compression is much less than the characteristic stiffness of the particles in contact. The linearized equations of motion for the finite system are the same as Eqs.~(\ref{linearizedEOM}), except the equations for the first and last particles which are given by the following expressions: 
\begin{equation}
\begin{aligned}
m\ddot{u}_{1} = F_{1}-\beta_1[u_{1}] - \beta_1[u_{1} - u_{2}], \\
m\ddot{u}_{21} = \beta_1[u_{20} - u_{21}]. \\
\label{linearizedEOMBC}
\end{aligned}
\end{equation}
where $F_1$ is the force applied to the first particle by the actuator.
Next, we apply the state-space approach, which can be written as \cite{Ogata}:
\begin{equation}
\begin{aligned}
\mathbf{\dot{x}}&=\mathbf{A}\mathbf{x}+\mathbf{B}\mathbf{u} \, \\
			\mathbf{y} &=\mathbf{C}\mathbf{x}+\mathbf{D}\mathbf{u}, \\
\label{statespace}
\end{aligned}  
\end{equation}
where vectors $\mathbf{x}$, $\mathbf{u}$, and $\mathbf{y}$ are the state, input, and output vectors, respectively.
Matrices $\mathbf{A}$, $\mathbf{B}$, $\mathbf{C}$, and $\mathbf{D}$ are called state, input, output and direct transmission matrices, respectively. 
We choose as an input to the system the force $F_{1}$, i.e., $\mathbf{u}=F_{1}$ and as an output $\mathbf{y}=F_N=\frac{\beta_1[u_{20} - u_{21}]}{2}$, the averaged force of the two contacts  of the last particle (which is analogous to what is measured by the embedded dynamic force sensor in our experimental setup)~\cite{NesterenkoSensor,dar05,dar05b,dar06}. 
Thus, for the linear system of Fig.~\ref{Fig1}(b), we obtain
\begin{equation*}
	\mathbf{x} =
\left(
\begin{array}{c}
u_1 \\
\vdots \\
u_{N-1} \\
u_N \\
\dot{u_1} \\
\vdots \\
\dot{u_N} \\
\end{array} \right)\,,
\end{equation*}
\begin{equation*}
	\mathbf{A} =
\left(
\begin{array}{c|c}
\mathbf{0} & ~\mathbf{I}  \\ \hline
\mathbf{M}^{-1}\mathbf{K}  & ~\mathbf{0}  \\
\end{array} \right)\,,
\end{equation*}
where,
\begin{equation*}
	\mathbf{M} =
\left(
\begin{array}{ccccccc}
m 			& 0 			& 0 		& \ldots 	& 0 			& 0			& 0\\
0 			& M 			& 0 		& \ldots 	& 0 			& 0			& 0\\
0 			& 0 			& m 		& \ldots 	& 0 			& 0			& 0\\
\vdots 	&\vdots 	&\vdots & \ddots 	&\vdots 	&\vdots &\vdots \\
0 			& 0 			& 0 		& \ldots 	& m 			& 0			& 0\\
0 			& 0 			& 0 		& \ldots 	& 0 			& M			& 0\\
0 			& 0 			& 0 		& \ldots 	& 0 			& 0			& m\\

\end{array} \right)\,,
\end{equation*}
\begin{widetext}
\begin{equation*}
	\mathbf{K} = 
	\left(
\begin{array}{ccccccccc}
-2\beta_1 			& \beta_1 				& 0 							& 0 							& 0 			& \ldots 					& 0 							& 0					& 0	\\
\beta_1 				& -2\beta_1 			& \beta_1 					& 0 							& 0 			& \ldots 					& 0 							& 0					& 0	\\
0 						& \beta_1				& -\beta_1-\beta_2	& \beta_2 					& 0 			& \ldots 					& 0 							& 0					& 0	\\
0							& 0							& \beta_2					& -\beta_2-\beta_1	& \beta_1  & 								& 0 							& 0     		& 0 \\
\vdots 				&\vdots 				&									&									& \ddots	&									&									&\vdots			&	\vdots	\\
0 						& 0			 				& 0 							& 			 					& \beta_1 	& -\beta_1-\beta_2 	& \beta_2 					& 0					& 0	\\
0 						& 0				 			& 0			 					& \ldots					& 0 			& \beta_2 					& -\beta_2-\beta_1 	& \beta_1		& 0	\\
0 						& 0							& 0								& \ldots					& 0 			& 0 							& \beta_1 					& -2\beta_1	& \beta_1	\\
0							& 0							& 0								& \ldots					& 0  			& 0								& 0 							& \beta_1    & -\beta_1 \\
\end{array} \right).\end{equation*}
\end{widetext} 
\begin{equation*}
	\mathbf{B} =
\left(
\begin{array}{c}
0 \\
\vdots \\
0 \\
1/m \\
0 \\
\vdots \\
0 \\
\end{array} \right)\,,
\end{equation*}
\begin{equation*}
	\mathbf{C} =
\left(
\begin{array}{ccccccc}
0 & \ldots  & \frac{\beta_1}{2} & -\frac{\beta_1}{2} & 0 &\ldots & 0 \\
\end{array} \right)\,,
\label{statespacesubmatrices}
\end{equation*}
and $\mathbf{D}=0$.
$\mathbf{0}$ is a zero matrix and $\mathbf{I}$ is the identity matrix (both of size $N \times N$).

\begin{figure}[h]
\begin{center}
\includegraphics[width=9cm,height=5cm]{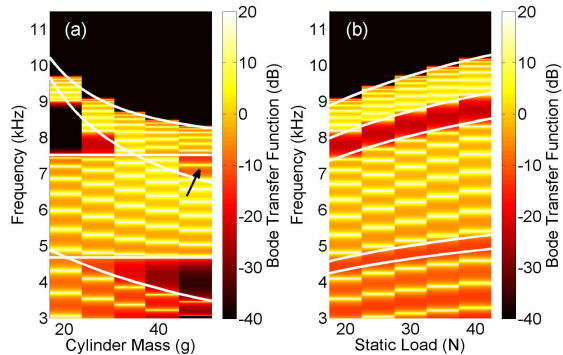}
\end{center} 
\caption{\label{Fig4}[Color online] Bode transfer function ($|H(i\omega)|$) for the experimental configurations:
(a) the five diatomic (three-particle unit cell) granular crystals with varied cylinder length for fixed $F_0$=20~N static compression, and (b) the fixed cylinder length $L=12.5$~mm ($M=27.3$~g) granular crystal with varied static load. Solid white lines are the cutoff frequencies calculated from the dispersion relation of the infinite system. The black arrow in (a) denotes the eigenfrequency of a locally supported defect mode.}
\end{figure}

We use the formulation in Eqs.~\ref{statespace} with MATLAB's (R2008b) $bode$ function to compute the bode diagram of the frequency response for the experimental configurations described. The bode diagram is the magnitude of the transfer function $H(s)=D+\mathbf{C}(s\mathbf{I}-\mathbf{A})^{-1}\mathbf{B}$, where $s=i\omega$~~\cite{Ogata}. 
We plot the bode transfer function $|H(i\omega)|$ for the five previously described diatomic (three-particle unit cell) chains with varied cylinder length for fixed $F_0$=20~N static compression, (Fig.~\ref{Fig4}(a)), 
and with varied
static compression ($F_0=[20,~25,~30,~35,~40]$~N) for fixed cylinder length $L=12.5$~mm ($M=27.3$~g) (Fig.~\ref{Fig4}(b)). 

We truncate the visualization in Fig.~\ref{Fig4} below $-40$~dB and above $20$~dB as a visual aid to maintain clarity of the frequency region of interest. This resembles experimental conditions, as the noise floor of our measurements is approximately $-38$~dB (as can be seen in Fig.~\ref{Fig5}) and the presence of dissipation in our experiments reduces the sharpness of the resonant peaks in contrast to those predicted by the state-space analysis.
Attenuating and propagating frequency regions for this formulation match well with the cutoff frequencies of the infinite system (see Eqs.~(\ref{cutofffrequencies})), denoted by the solid lines plotted in Fig.~\ref{Fig4}. The high amplitude (bright) peaks correspond to the eigenfrequencies of the system, the modes of which are spatially extended. However, for certain cylinder lengths, we also observe an eigenfrequency located in the second gap of the linear spectrum (denoted by an arrow in Fig.~\ref{Fig4}(a)). These modes result from the break in periodicity due to the presence of the actuator ``wall" (acting like a defect in the system). In our setup, it can be seen in Fig.~\ref{Fig1}(b) that the first particle (which is spherical) is coupled to both its nearest neigbors via springs characterized by spherical-planar contact ($\beta_1$). This is unique within the chain and forms a type of locally supported defect mode. When the frequency of this mode lies within a band gap the mode becomes spatially localized around the first particle and its amplitude decays exponentially into the chain. Furthermore, as our chains are relatively short and the gap that the localized modes occupy relatively narrow (in frequency), the spatial profile is found to be almost similar to the extended modes. This suggests that it may be experimentally difficult to differentiate these modes from their extended counterparts.

\section{\label{sec:level1}Experimental linear spectrum}

\begin{figure}[h]
\begin{center}
\includegraphics[width=9cm,height=5cm]{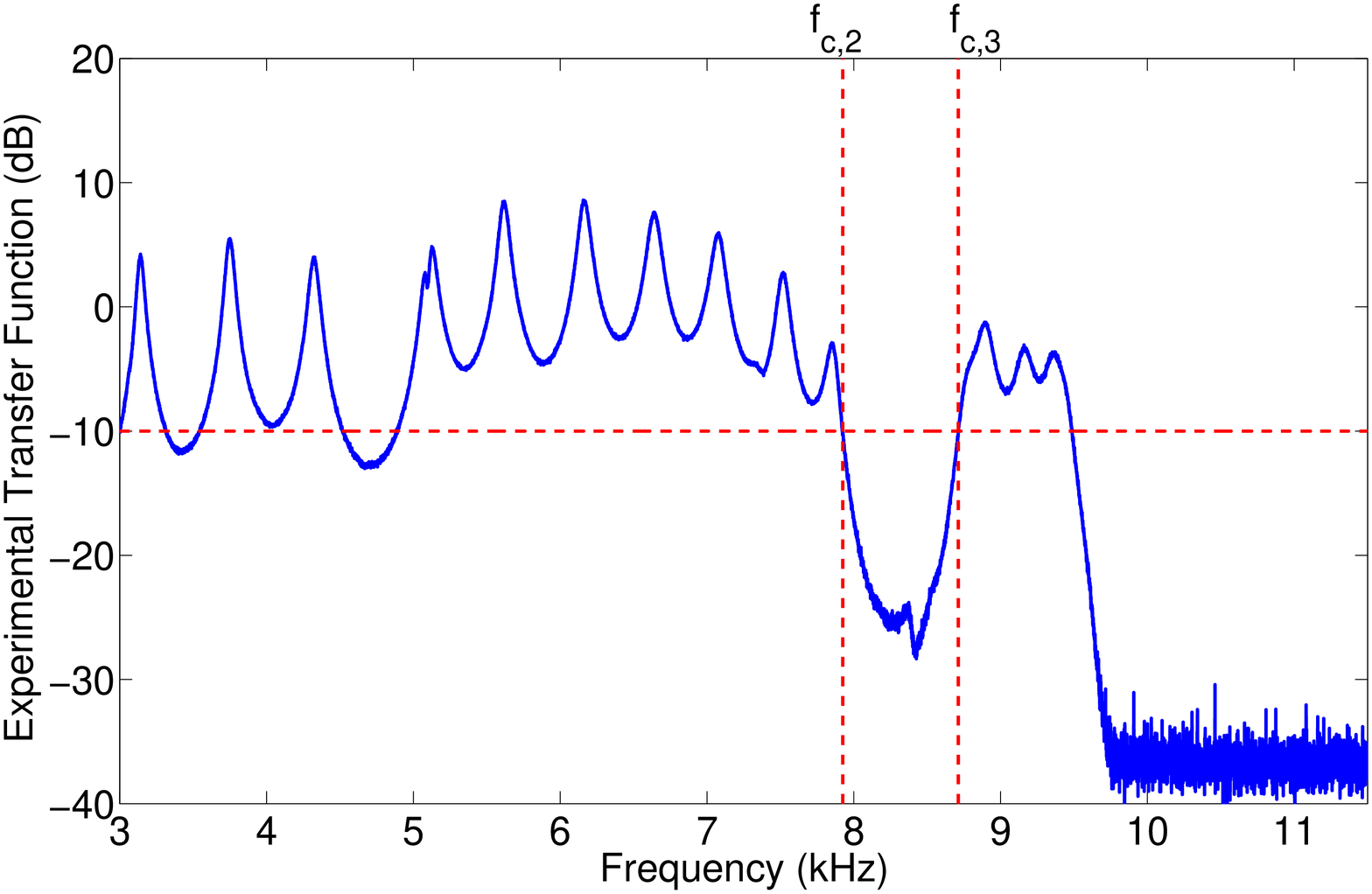}
\end{center} 
\caption{\label{Fig5}[Color online] Experimental transfer function for the $L=12.5$~mm ($M=27.3$~g), $F_0=20$~N granular crystal. The horizontal dashed line is the $-10$~dB level used to experimentally determine the $f_{c,2}$ and $f_{c,3}$ band edges which are denoted by the vertical dashed lines}
\end{figure}

\begin{figure}[h]
\begin{center}
\includegraphics[width=9cm,height=5cm]{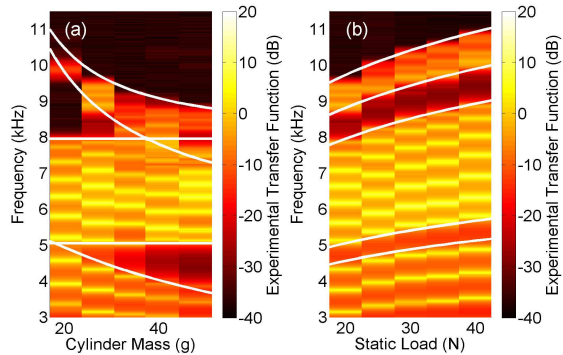}
\end{center} 
\caption{\label{Fig6}[Color online]  Experimental PSD transfer functions for the experimental configurations described in Fig.~\ref{Fig4}: (a) the five diatomic (three-particle unit cell) granular crystals with varied cylinder length for fixed $F_0$=20~N static compression, and (b) the fixed cylinder length $L=12.5$~mm ($M=27.3$~g) granular crystal with varied static load. Solid white lines are the cutoff frequencies from the dispersion relation using experimentally determined Hertz contact coefficients $A_{1,exp}$ and $A_{2,exp}$.}
\end{figure}

We experimentally characterize the linear spectrum of the previously 
described diatomic chains with sphere-cylinder-sphere unit cells 
for varied cylinder length and static load. We apply a low-amplitude (approximately $200$~mN peak) bandwidth limited ($3-15$~kHz) noise excitation with the piezoelectric actuator. We measure the dynamic force using a sensor embedded in the last particle of the granular crystal as shown in Fig.~\ref{Fig1}. We compute the power spectral density (PSD \cite{PSD}) of the measured dynamic force history over $1.3$~s intervals, and average the PSD over $16$ acquisitions. We normalize the averaged PSD spectrum by the average PSD level in the $3-7.5$~kHz range of the $L=12.5$~mm ($M=27.3$~g), $F_0=20$~N granular crystal response to obtain the transfer functions shown in Fig.~\ref{Fig5} and Fig.~\ref{Fig6}. More specifically, Fig.~\ref{Fig5} shows the experimental transfer function in more detail for the $L=12.5$~mm ($M=27.3$~g), $F_0=20$~N granular crystal.  

As in \cite{Boechler10}, we observe that the experimentally determined spectra are upshifted in frequency from the theoretically derived spectra for all configurations tested. Because of this we use the measured spectra to extract
 the effective elastic properties of our system. For the $F_0=[20,~25,~30,~35,~40]$~N, fixed cylinder length $L=12.5$~mm ($M=27.3$~g) granular crystals, we measure the frequencies of the $-10$~dB level of the PSD transfer function corresponding to the second band gap ($f_{c,2}$ and $f_{c,3}$). We use these experimentally determined frequencies to solve for two experimentally determined Hertzian contact coefficients of our system $A_{1,exp}$ and $A_{2,exp}$ using the previously described equations for $A_1$,~$A_2$,~$\beta_1$,~$\beta_2$,~$f_{c,2}$, and $f_{c,3}$. An example of the determination of $f_{c,2}$ and $f_{c,3}$ for the $L=12.5$~mm ($M=27.3$~g), $F_0=20$~N granular crystal is shown in Fig.~\ref{Fig5}. We compare the experimentally determined $A_{1,exp}$ and $A_{2,exp}$ to the theoretically determined $A_{1}$ and $A_{2}$ in Table~\ref{table_th_vs_exp_linear} (error ranges indicate the standard deviation resulting from the measurements at the five different static loads). 
\begin{table}[h]
\begin{tabular}{|c|c|c|}
\hline       & $A_1$ [N/$\mu$m$^{3/2}$] 		& $A_2$ [N/$\mu$m$^{3/2}$] 	\\ 
\hline Theory   & $14.30$            						& $10.11$                 		\\ 
\hline Experiments  & $18.04 \pm 0.44$ 						& $11.48 \pm 0.06$      		\\
\hline 
\end{tabular}
\caption{\label{table_th_vs_exp_linear} Hertz contact coefficients derived from standard specifications~\cite{ElasticProperties} ($A_1$ and $A_2$) versus
coefficients derived from the measured frequency cutoffs ($A_1$,exp and $A_2$,exp), for the ($F_0=[20,~25,~30,~35,~40]$~N) fixed cylinder length $L=12.5$~mm ($M=27.3$~g) granular crystals.}
\end{table}
As the equations for the  five non-zero cutoff frequencies (see Eqs.~\ref{cutofffrequencies}) in our granular crystals are dependent on some combination of $A_1$ and $A_2$, the choice of using $f_{c,2}$ and $f_{c,3}$ to solve for $A_1$ and $A_2$ is not unique and other combinations of cutoff frequencies could be used similarly. 

In previous work \cite{Boechler10}, numerous possible explanations for the upshift in the spectrum have been identified. Uncertainty in the standard values of material parameters~\cite{ElasticProperties} or deviations in the local radius of curvature due to surface roughness could result in the material behaving more stiffly~\cite{Coste1999}. Deviations from Hertzian behavior could potentially be caused by the dynamic loading conditions, non-Hookean elastic dynamics (due to nonlinear elasticity or plasticity), or dissipative mechanisms (viscoelasticity, solid friction), and could result in a shift in the exponent $p$ or in the effective contact coeffcient $A$~~\cite{Coste1999,Job2005,Carretero2009}. We also observe that the contact coefficient $A$ between the cylindrical and spherical particles has the larger deviation from theory. This deviation could be attributed mainly to the cylindrical particles, due to characteristics not shared by the spherical particles. Such characteristics could include: surface roughness particular to the manufacturing process of the cylindrical particles, or plastic deformation occuring closer to the surface as compared to spherical particles.  

In Fig.~\ref{Fig6}, we plot the experimentally determined PSD transfer functions for the five previously described diatomic (three-particle unit cell) chains with varied cylinder length for fixed $F_0$=20~N static compression (Fig.~\ref{Fig6}(a)), and static compression $F_0=[20,~25,~30,~35,~40]$~N, for fixed cylinder length $L=12.5$~mm ($M=27.3$~g) (Fig.~\ref{Fig6}(b)). We plot with solid white lines the cutoff frequencies from the dispersion relation calculated using the experimentally determined Hertz contact coefficients $A_{1,exp}$ and $A_{2,exp}$. We observe good agreement between the semi-analytically 
derived cutoffs (i.e, from the theoretical dispersion relation but 
using $A_{1,exp}$ and $A_{2,exp}$) and the experimental spectra. By comparing Fig.~\ref{Fig6} to Fig.~\ref{Fig4} we observe good qualitiative agreement between the numerical (state-space) and experimental spectra. Comparing the experimentally and theoretically determined cutoff frequencies, we observe an average (over all experimental configurations) upshift in the experimental frequency cutoffs versus the theoretically determined frequency cutoffs of: $5.8\%$ in $f_{c,2}$, $8.1\%$ in $f_{c,3}$, $8.1\%$ in $f_{c,4}$, $5.4\%$ in $f_{c,5}$, and $7.0\%$ in $f_{c,6}$.

The demonstrated attenuation of the elastic wave propagation in frequency regions lying within the band gaps of the granular crystals shows that such systems have potential for use in a wide array of vibration filtering applications. Furthermore, the tunability displayed (achievable from material selection, shape, size, periodicity
and application of static compression) offers significant 
potential for attenuating a wide spectrum of undesired frequencies.  
\newline

\section{\label{sec:level1}Conclusions}

In this work, we describe the tunable vibration filtering properties of a 1D granular crystal composed of periodic arrays of three-particle unit cells. The unit cells are assembled with elastic beads and cylinders that interact via Hertzian contact. Static compression is applied to linearize the dynamics of particles interaction and to tune the frequency ranges supported by the crystal. We measure the transfer functions of the crystals using state-space analysis and experiments, and we compare the results with the corresponding theoretical dispersion relations. Up to three distinct pass bands and three (two finite) band gaps are shown to exist for selected particle configurations. The tunability of the band edges in the crystal's dispersion relation is demonstrated by varying the applied static load and the cylinder length.

In the present work, we restrict our considerations to the study of near linear, small amplitude excitations. A natural extension of this work would involve the
examination of nonlinear excitations within the bandgaps of such
granular chains \cite{Boechler10}. In particular,
it would be relevant to compare the properties of localized 
nonlinear waveforms in different gaps of the linear spectrum.
Such studies will be reported in future publications.
\newline
\newline

\begin{acknowledgments}
We thank St\'{e}phane Job for help with the experimental setup. PGK gratefully
acknowledges support from NSF-CMMI-1000337. GT and PGK acknowledge support from the A.S. Onassis
Public Benefit Foundation through Grant  RZG 003/2010-2011. CD Acknowledges support from NSF-CMMI-0844540 (CAREER) 
and NSF-CMMI-0969541.
\end{acknowledgments}


\end{document}